\documentclass[conference]{IEEEtran}
\IEEEoverridecommandlockouts
\usepackage{graphicx} 
\usepackage{amsmath}  
\usepackage{amssymb}
\usepackage{color}
\usepackage{multirow}
\usepackage{flushend}

\title{Error Checking for Sparse Systolic Tensor Arrays}
\author{%
\IEEEauthorblockN{Christodoulos Peltekis, Dionysios Filippas and Giorgos Dimitrakopoulos}
\IEEEauthorblockA{\small Electrical and Computer Engineering, Democritus University of Thrace, Xanthi, Greece\thanks{This work was supported by a research grant of Siemens EDA to Democritus University of Thrace for ``HLS Research for Systems-on-Chip''.}}}

\begin{document}

\maketitle

\begin{abstract}
Structured sparsity is an efficient way to prune the complexity of modern Machine Learning (ML) applications and to simplify the handling of sparse data in hardware. In such cases, the acceleration of structured-sparse ML models is handled by sparse systolic tensor arrays. The increasing prevalence of ML in safety-critical systems requires enhancing the sparse tensor arrays with online error detection for managing random hardware failures.
Algorithm-based fault tolerance has been proposed as a low-cost mechanism to check online the result of computations against random hardware failures. In this work, we address a key architectural challenge with structured-sparse tensor arrays: how to provide online error checking for a range of structured sparsity levels while maintaining high utilization of the hardware.
Experimental results highlight the minimum hardware overhead incurred by the proposed checking logic and its error detection properties after injecting random hardware faults on sparse tensor arrays that execute layers of ResNet50 CNN.
\end{abstract}

\section{Introduction}
Machine Learning (ML) has achieved unprecedented success in various domains. 
To minimize memory storage and computation cost, ML model weights are pruned, resulting in sparse models~\cite{hoefler2021sparsity}. The derived zero weights are not stored and the corresponding computation is skipped. 

\begin{figure}[t]
\centering
\includegraphics[width=0.98\columnwidth]{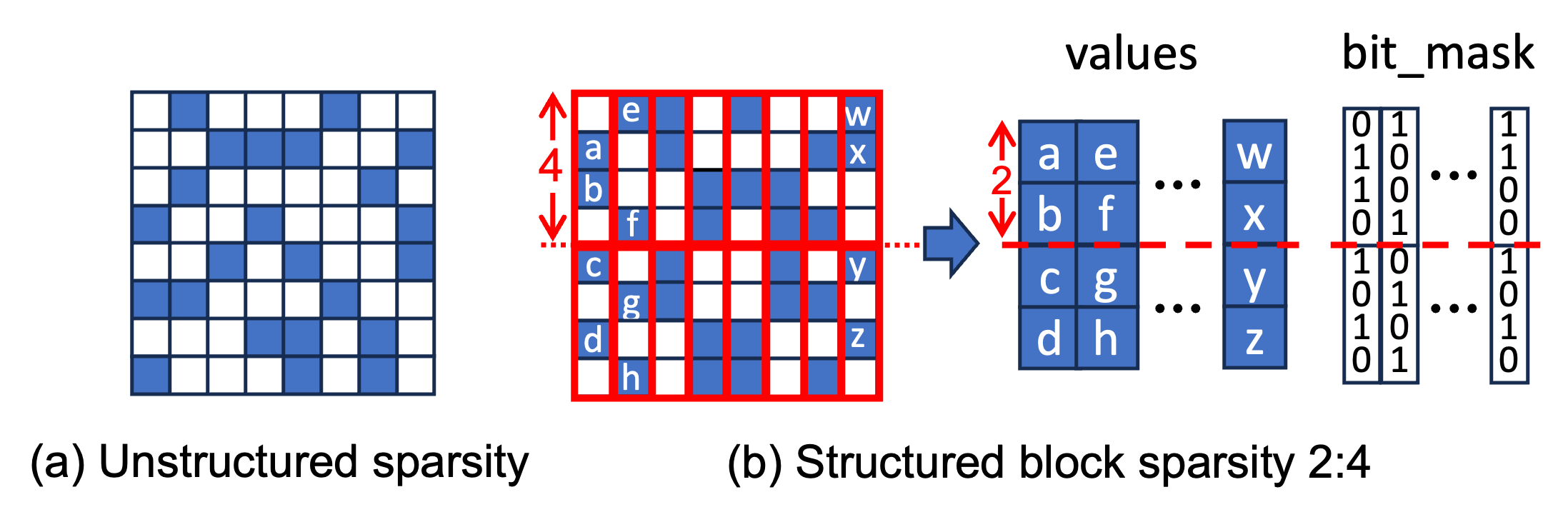}
\caption{Example of (a) unstructured sparsity; and (b) structured block sparsity of 2:4 (i.e., up to 2 non-zero elements in every 4 consecutive elements) and their respective packed storage with their associated bit masks.}
\label{f:unstructered-block-sparse}
\end{figure}

The achieved sparsity can either be \textit{unstructured}~\cite{rigl}, or \textit{structured}~\cite{nvidia-block-sparse,learning-n-m}. In unstructured sparsity, as shown in Fig.~\ref{f:unstructered-block-sparse}(a), there is no constraint on the locations of the zeros and multiple indexes should accompany the non-zero data. On the contrary, in $N$:$M$ structured sparsity, a fixed number of $N$
non-zero elements may be present within a column block of $M$ rows. For instance, in Fig.~\ref{f:unstructered-block-sparse}(b), for every four elements in each column, there are up to two non-zero elements (i.e., 2:4 sparsity). In practice, the size of blocks $M$ is small and indexing of non-zero elements inside each block is done with bit masks.

The acceleration of ML models relies primarily on equivalent matrix multiplications that inherently map to systolic arrays (SAs)~\cite{why-systolic, scalesim}. In case of structured sparse data, computation is performed by lower-cost sparse systolic tensor arrays~\cite{s2ta, vegeta}. Dense SAs are built from scalar PEs and sparse tensor arrays from tensor PEs. A scalar PE accepts an operand per cycle and computes a single multiply-add with the locally-stored weight. On the contrary, a tensor PE accepts a fixed-size block of elements per cycle~\cite{s2ta}. Inside each tensor PE, the operands of each block that should be multiplied with the locally stored weights are selected dynamically by matching the indexes of the corresponding non-zero data. Besides this structural difference, sparse tensor arrays can implement any of the dataflows supported by dense SAs.

Checking the correctness of sparse matrix multiplications in the presence of random hardware faults is necessary for safety in critical applications~\cite{standard}. Recent works~\cite{junior2022reliability, kundu2021toward, chitty2019impact} quantified the robustness and reliability of systolic accelerators when affected by random faults. 
Also, new fault-tolerant architectures were proposed that bypass or deactivate faulty PEs supported also by appropriate weight pruning and model compression~\cite{zhao2022fsa, chitty2020model, zhang2019fault, lee2023strait}. In such cases, the fault detection was done with application-specific test vectors~\cite{vacca2023runsafer}.

Managing random hardware faults~\cite{soft-errors, unreliable}, requires special hardware modules for fault detection~\cite{fault-tolerant-systems}. Faults should be detected \emph{online} and possibly within a few cycles of their occurrence, thus simplifying recovery. 

Algorithm-Based Fault Tolerance (ABFT)~\cite{abft,abft-fft, fault-tolerance-sa} offers a low-cost mechanism to detect errors in matrix-based computations~\cite{abft-pratical, conv-checksum-tvlsi} by comparing the true output checksum with a predicted one. For dense SAs, specialized checksum circuits have been proposed~\cite{low-voltage-sa, light-abft, online-date23} that compute checksums and perform error detection, while the systolic array processes input data. 

In this work, our goal is to leverage the ABFT techniques already known for dense SAs and design for the first time --to the best of our knowledge-- efficient checksum computation architectures for sparse tensor arrays. The contributions of this work can be summarized as follows:
\begin{itemize}
\item 
Sparse systolic tensor arrays are enhanced with checksum logic that allow checking the output of matrix multiplication a few cycles after finishing the actual computation. Checksum logic is inserted at the borders of the sparse tensor array without altering its functionality or imposing any other constraint to its design.
\item
ABFT checksum logic, as the proposed one, requires accumulators for adding input or output results. Thus, inevitably, the minimum required bitwidth of integer accumulators exceeds that of incoming data. To handle the wider-width operations needed for computing checksums, we perform checksum checking in a digit-serial fashion. In this way, checksum logic can operate on arbitrary large input matrices.
\item 
Experimental results quantify both the area and power overhead of the proposed error checking mechanism when applied in state-of-the-art sparse tensor arrays as well its fault detection performance. In all examined cases, the error checking introduces only a marginal overhead of less than 5\% and offers fault detection above 90\% for realistic scenarios.
\end{itemize}

\section{Background on Error checking on dense systolic arrays}

\begin{figure}[h!]
\centering
\includegraphics[width=0.85\columnwidth]{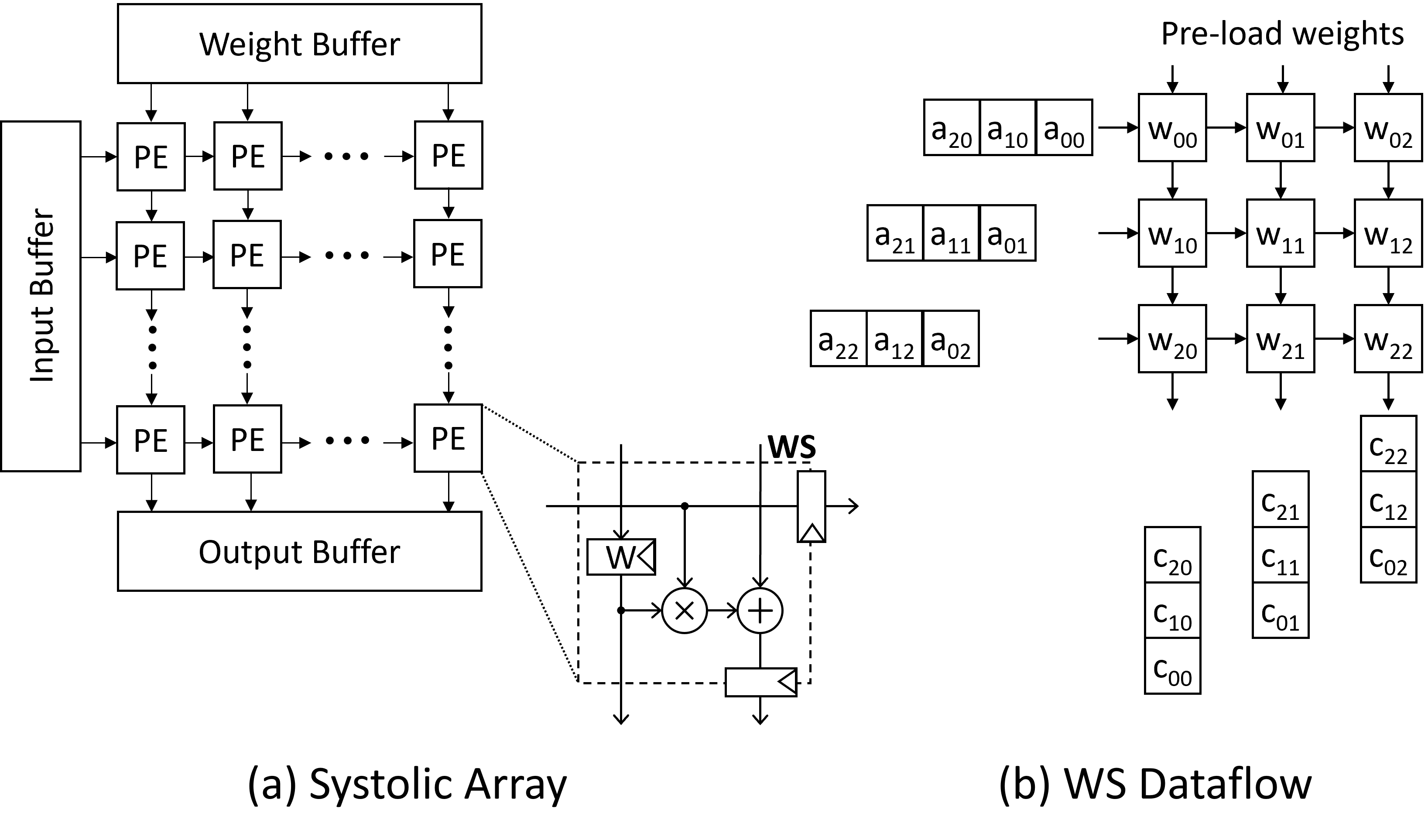}
\caption{A typical systolic array for a fully dense matrix multiplication following the weight-stationary (WS) dataflow.}
\label{f:dense-sa}
 \end{figure}

The SA hardware structure typically comprises an array of Processing Elements (PEs), as illustrated in Fig.~\ref{f:dense-sa}. 
Each PE consists of a multiplier and an adder and necessary registers to appropriately pipeline the streaming operation. 
Local memory banks positioned at the West and North edges of the array supply the SA with input data, while the resulting output is gathered on the South. For matrices exceeding the SA's dimensions, matrix multiplication is performed in tiles, with each tile's size aligning with that of the SA.

The SA's chosen dataflow type shapes the internal structure of the PEs and determines the execution of matrix multiplication.
Weight-stationary dataflow implementations, in particular, involve pre-loading matrix $W$ (the 'weights') into the SA, with matrix $A$ (the 'inputs') being supplied from the West side. The outcomes are collected at the South end of each column. This arrangement maintains matrix $W$ stationary.

Following ABFT methodology~\cite{abft}, to check the result of matrix multiplication 
$C = A\, W$, we need to compare the \emph{actual checksum} of all output elements of $C$ with a \emph{predicted checksum}.  In~\cite{abft} it was shown that output checksum can be predicted by the dot product of the per-column checksum vector of input $A$ and the per-row checksum vector of weights $W$. The column checksums of $A$ and the row checksums of $W$ can be computed either offline or online~\cite{low-voltage-sa, online-date23}. Then, the required dot product between the two checksum vectors can be computed using additional multiply-add units or by reusing the operators of the SA.

\section{Online Error Checking for Sparse Systolic Tensor Arrays}
A sparse tensor array computes the matrix product $C\!=\! A\, W$ of a dense input matrix $A$ and a structured-sparse matrix $W$. It follows the same systolic array archetype but it is composed of tensor PEs (TPEs) instead of the scalar PEs, as shown in Fig.~\ref{f:dense-sa}. An example of a two-row sparse tensor array that can be configured for 2:4 and 1:4 sparsity is shown in Fig.~\ref{f:sparse-arch}.

Each TPE consists of two registers that store the two non-zero elements found every four elements of the pruned weight matrix. These registers are loaded with the appropriate weights during a separate weight-loading phase exactly as in a typical weight-stationary dataflow. 

Each TPE accepts four input values from its west side that belong to the same row of $A$. To perform matrix multiplication two of the four input elements will be selected and multiplied with the locally-stored weights. The selection of the appropriate pair of inputs is based on the column indexes of the pair of stored weights and it is performed by the two 4-to-1 multiplexers. The computed products are added in each column of the tensor array. When structured sparsity of 1:4 is enabled, only one of the two multiplexers and multipliers per TPE will be activated. This process continues for all incoming rows of $A$, e.g., different rows of $A$ arrive in every TPE in blocks of four elements and the stationary weights select which ones should be used for computation.

\subsection{ABFT on Sparse Tensor Arrays}
Following the ABFT methodology, to predict the checksum of the multiplication we have to compute the dot product of the per-column checksum vector of $A$ and the per-row checksum vector of $W$. This predicted checksum should be compared to the actual checksum of matrix multiplication.

\begin{figure}
\centering
\includegraphics[width=0.98\columnwidth]{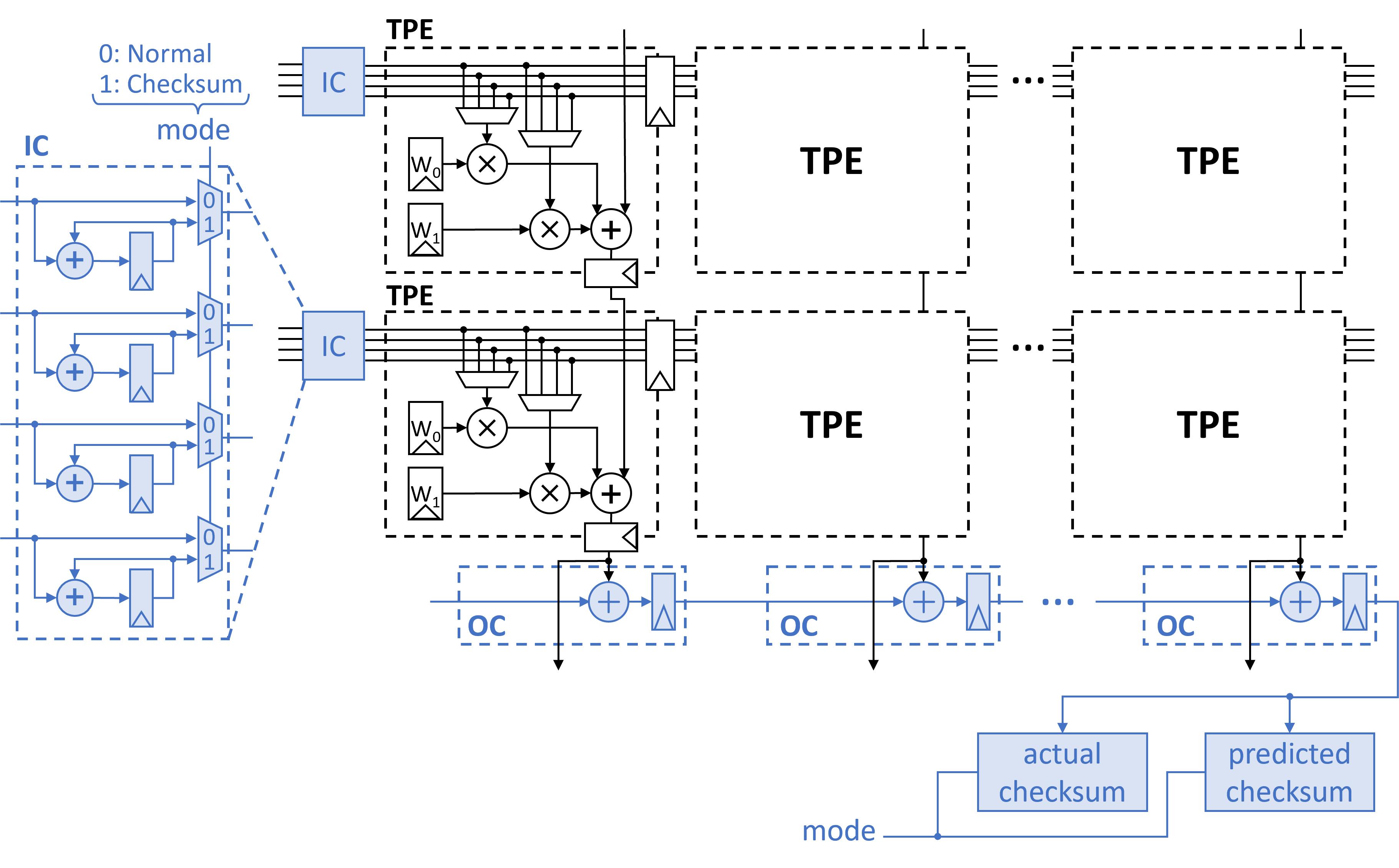}
\caption{A two-row sparse systolic tensor array that can be configured for $2$:$4$ and $1$:$4$ sparsity patterns. ABFT error checking that computes the actual and the predicted checksum is placed at the periphery of the array.}
\label{f:sparse-arch}
\end{figure}

The actual checksum is computed in the OC modules put on the south edge of the sparse tensor array, as shown in Fig.~\ref{f:sparse-arch}. The OC modules add in a pipelined manner the results produced at each column of the sparse tensor array. The total sum is then accumulated at the actual-checksum accumulator shown in the south-east corner of Fig.~\ref{f:sparse-arch}.

In this work, computing the predicted checksum utilizes both the sparse tensor array as well as extra checking logic. First, we need to compute the per-column checksum vector of $A$. Similar to~\cite{low-voltage-sa, online-date23}, this is computed online using the input accumulator blocks (IC) attached at the west side of the sparse tensor array. Each IC block, shown in Fig.~\ref{f:sparse-arch}, consists of four accumulators. This requirement stems from the fact that we do not know beforehand which inputs will be selected by the weights stored in a row of the TPE. Thus, we need to accumulate all incoming inputs of dense matrix $A$. 

Then, we need to compute the dot product of the per-column checksum vector of $A$ and the per-row checksum vector of $W$. To do this, in this work, we re-use the TPEs of the sparse tensor array.
For instance, after the last row of matrix $A$ enters the sparse tensor array, we gradually stop normal computation and push inside the sparse tensor array the column checksum vector $\sum_i A_{i,j}$ accumulated at the IC blocks using the same skewed data arrival pattern. Effectively, this action extends the computation with an ``extra row'' of checksum inputs.
This column checksum vector of $A$ will propagate inside the sparse tensor array and be multiplied with the corresponding non-zero weights. At the bottom of the $k$th column of the sparse tensor array the sum $\sum_j (\sum_i A_{i,j}) W_{j, k}$ will be produced. This sum is equivalent to $(\sum_i A_{i,j})(\sum_j W_{j, k})$ that was originally sought.
In this way, computing separately the per-row checksum vector of sparse matrix $W$ (i.e., $\sum_j W_{j, k}$) is redundant and can be skipped. Also, the same non-zero elements of $W$ that affect the result of matrix multiplication contribute to the computed checksum.

Checksum prediction is completed at the south edge of the sparse tensor array. The checksums produced per column after normal computation has finished should be added to complete the predicted checksum. This addition is also done by the OC modules. 
However, in this case, since each sum computed by an OC module refers to the predicted checksum and not the actual one, it gets accumulated to a predicted-checksum accumulator at the south-east corner.

\subsection{Tackling the increased width of accumulators with digit-serial additions}
To save power and to reduce the memory footprint of ML models, model pruning and quantization is applied during training, while inference is executed 
using integer arithmetic. Without loss of generality, let's assume that the sparse tensor array operates on $8$-bit quantized inputs and weights producing per column 24-bit results. 

The proposed ABFT checking logic computes the predicted checksum by reusing the multiply-add logic of the sparse tensor array. In particular, in the last wave of computation, the sparse tensor array does not receive the data of input matrix $A$ but their per-column checksum. However, the bit width of the accumulators is much larger than the input bit width, e.g., accumulating 256 8-bit inputs we would need 16-bit accumulators. Therefore, passing this 16-bit accumulated value through the 8-bit inputs of the sparse tensor array cannot be done in the same cycle.

\begin{figure}[t]
\centering
\includegraphics[width=0.92\columnwidth]{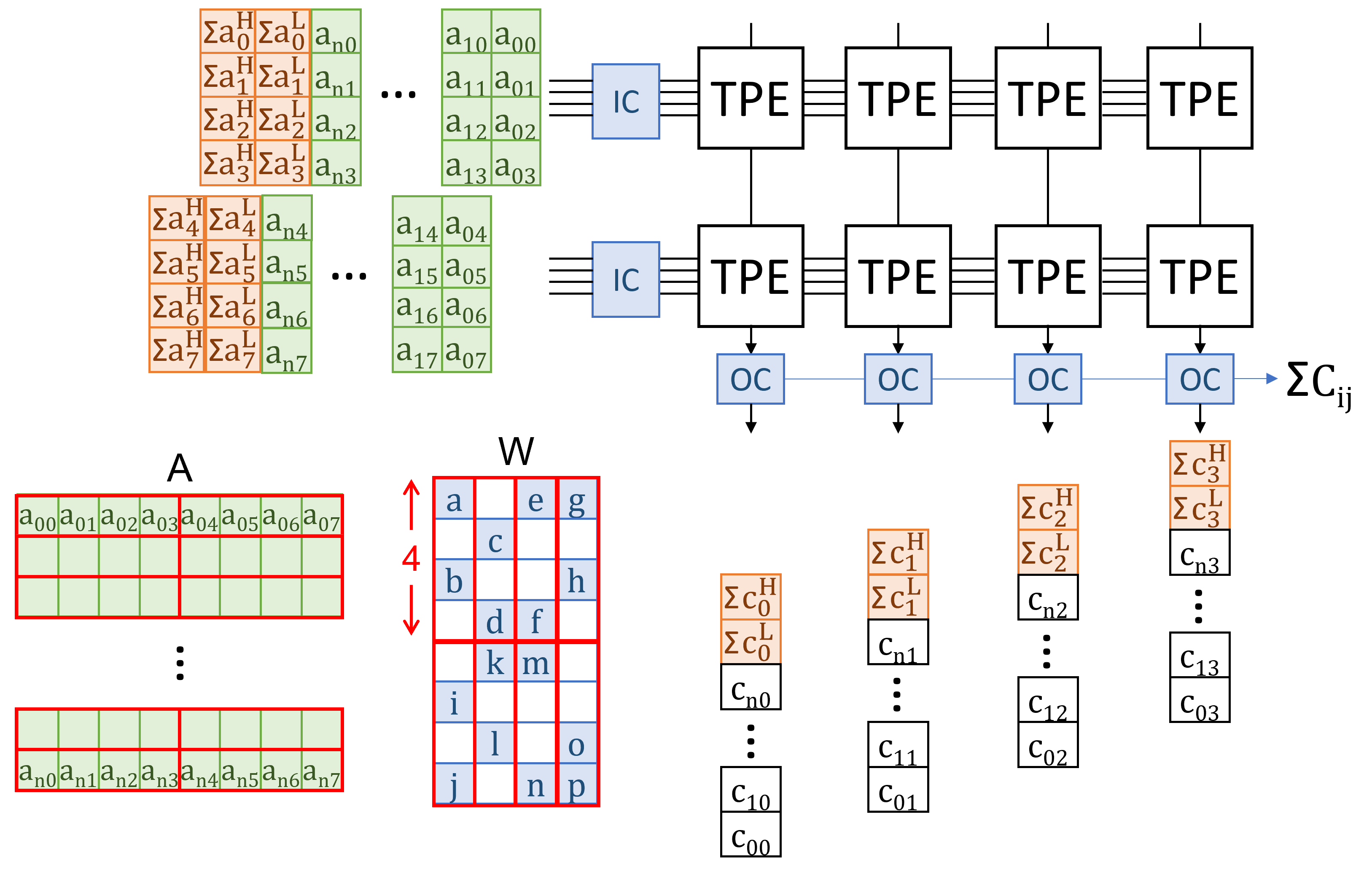}
\caption{Checksum computation follows back-to-back normal computation. Checksums are computed in a digit-serial manner reusing the fixed structured of the sparse tensor array.}
\label{f:sparse-ws-dataflow}
\end{figure}

To address this issue, we perform the \emph{checksum operation} for the first time --to the best of our knowledge-- in a \emph{digit-serial manner}. The input checksum is broken into smaller 8-bit wide digits and are fed into the SA one digit per cycle. 
The 8-bit checksum digits are propagated inside the sparse tensor array and produce new checksum values at the bottom of each column of the array. The derived digit-wise checksums are accumulated using the OC modules and the predicted-checksum accumulator. This accumulator handles also the needed shifting of the arriving data and their sign extension to perform the signed digit-serial addition properly. The back-to-back evolution of normal and checksum computation, where checksum is computed in a digit-serial manner is highlighted in Fig.~\ref{f:sparse-ws-dataflow}.

In the general case, assuming that the input accumulators are by construction $M$-bits wide and that the sparse tensor array operates on $N$-bit quantized integers, the maximum number of rows of $A$ that can be safely accumulated is $t = 2^M/2^N$. Thus, every $t$ cycles, we need to interrupt normal operation and compute the respective checksum. Digit-serial checksum computation involves $M/N$ digits that are passed serially one after the other to the sparse tensor array. In this way, we can support large input matrices and 
large per-column checksum values without altering the design of the sparse tensor array. Digit-serial checksum computation can be equally applied to dense systolic arrays.

\section{Experimental results}
In the experimental results, we aim to highlight two aspects of the proposed architecture. In the first set of experiments, we measure the hardware overhead induced by the error checking logic in terms of area and power to the sparse systolic tensor array. 
In the second set of experiments, we explore the fault detection properties of the checker, and the impact of sparsity to its performance.

\subsection{Hardware complexity evaluation}
The examined sparse tensor arrays and their associated error checking logic have been been synthesized to Verilog RTL using Catapult HLS. 
Clock frequency target was set to 500 MHz.
The reported timing/area results were derived from the Oasys-RTL logic synthesis tool driven by a 45-nm standard-cell library. 
Power estimation was done using PowerPro tool by running inference in state-of-the-art CNN applications, such as Resnet50~\cite{resnet}, with inputs from ImageNet. The weights used in the experiments assume structured sparsity of $1$:$4$ and $2$:$4$ derived with Tensorflow for ResNet50~\cite{resnet}.

Each sparse tensor array follows the organization shown in Fig.~\ref{f:sparse-arch}. It consists of $R$ rows and $C$ columns of TPEs and can be configured for $2$:$4$ or $1$:$4$ structured sparsity. Effectively the sparse tensor array receives $4R$ inputs at each west edge TPE. Inputs and weights are 8-bit quantized integers. The necessary data width
of the output of each column of the sparse tensor arrays depends on the number of rows of the array. To have a uniform design for all examined cases, we assume that the output bitwidth equals 24 bits.

The checker consists of $R$ IC modules in the west edge of the sparse tensor array, $C$ OC modules on the south edge and two accumulators for the actual and the predicted checksum on the south-east edge. The input accumulators in IC are 16-bits wide, the adders and pipeline registers in the OC modules are 24 bits, while the actual and the predicted checksum accumulators are 48-bits wide.

\begin{figure}
\centering
\includegraphics[width=0.9\columnwidth]{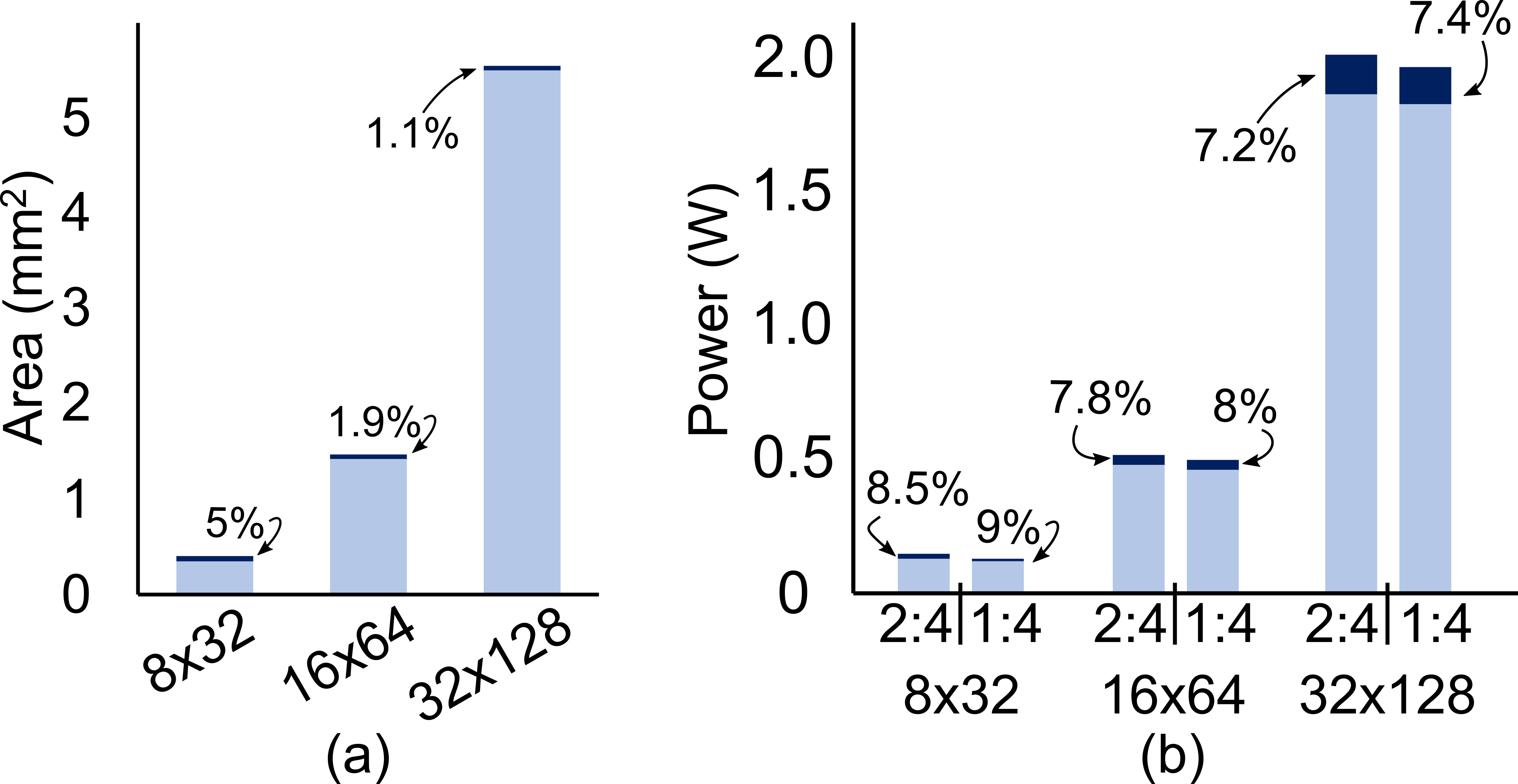}
\caption{The total (a) area and (b) power of sparse tensor arrays for various sizes and for 2:4 and 1:4 sparsity patterns. Both diagrams highlight separately the contribution of the checker's logic.}
\label{f:hw_complexity}
\end{figure}

The hardware complexity of the examined designs is summarized in Fig.~\ref{f:hw_complexity}. 
Fig.~\ref{f:hw_complexity}(a) presents the area overhead that is introduced by the checker for different sparse tensor array sizes.
For an $8\times 32$ sparse tensor array, the area overhead is small, around 5\%. This overhead gets further reduced to 1.9\% and 1.1\% as we increase the size of the sparse tensor array to $16\times 64$ and $32\times 128$, respectively.

Fig~\ref{f:hw_complexity}(b) reports the average power per layer of Resnet50 using inputs from ImageNet and weights that have a structured sparsity of $2$:$4$ and $1$:$4$. In the case of $2$:$4$ sparsity, the power overhead of the checker is 7.5\% on average for the three examined sparse tensor array sizes.
By changing the structured sparsity of the weights to $1$:$4$, the average power consumption is reduced. This is the result of the increased number of non-zero weights that leads to a higher inactivity of the sparse tensor array. On the other hand, since the number of IC and OC modules of the checker are not affected by the sparsity pattern, and is only determined by the size of the sparse tensor array, their power consumption remains almost constant. In this case, the power overhead ranges from 7.4\% to 9\% for the different sparse tensor array sizes. 

\subsection{Evaluation of fault detection properties}
For evaluating the quality of fault-detection of the proposed design we inject bit-flips in random clock cycles. Faults are injected to randomly selected storage elements covering both the registers of the sparse tensor array and the registers of the checking logic. 
Whether a fault will be injected on the sparse tensor array or the checker depends on the amount of their storage elements. A fault is more probable to hit the sparse tensor array than the checker's logic since it includes significantly more storage elements including the locally-stored weights as well as the pipeline registers placed at the borders of each TPE. 
The weights used during each fault-injection campaign correspond to \emph{all} pruned CNN layers of Resnet50 used also for power estimation.  

At the end of each matrix multiplication that corresponds to a CNN layer, we record the outcome of the fault injection campaign. The observed behavior may fall into one of four categories:
\begin{enumerate}
\item 
Detected: The checker detected an actual fault that occurred in the sparse tensor array.
\item 
Silent: The checker did not detect an actual fault that occurred in the sparse tensor array. The fault was masked at the checksum level although the checker remained fault free.
\item 
False positive: The checker flagged a fault detection, but no fault occurred in the sparse tensor array.
\item 
False negative: A fault occurred in the sparse tensor array and the checker did not detect it. Equivalently, the fault remained silent due an error inside the checker.
\end{enumerate}

\begin{table}[t]
    \caption{Fault Detection of the proposed error checking logic on a 8$\times$32 sparse tensor array that can be configured for 2:4 and 1:4 sparsity patterns for Resnet50 CNN layers.}
    \centering
    \begin{tabular}{|c||c|c|c|c|}
    \hline
        \# of faults   & \multicolumn{2}{c|}{1 fault} & \multicolumn{2}{c|}{1--
        5 faults} \\ 
    \hline
        Block Sparsity & $2$:$4$ & $1$:$4$ & $2$:$4$ & $1$:$4$  \\
    \hline\hline
        Detected       & 81.11\% & 74.15\% & 95.61\% & 93.34\%  \\
        Silent         & 12.46\% & 20.75\% & 2.15\%  & 4.95\%   \\
        False Positive & 2.36\%  & 2.58\%  & 0.58\%  & 0.69\%   \\
        False Negative & 4.07\%  & 2.51\%  & 1.65\%  & 1.02\%   \\
    \hline
    \end{tabular}
    
    \label{t:faults}
\end{table}

Table~\ref{t:faults} summarizes the outcome of 15000 independent fault-injection campaigns done on all pruned CNN Layers of Resnet50 (i.e., 300 independent fault-injection campaigns per CNN layer). Each fault injection campaign injects a different number of faults to a 8$\times$32 sparse tensor array and its checker. The same experiment is performed for 1:4 and 2:4 sparsity patterns.

In the case of a single fault injected, the majority of the faults belong to the detected and silent categories. Silent faults can occur in many scenarios. For sparse tensor arrays the most common case refers to faults affecting pipeline registers in the horizontal dataflow that refer to inputs that are not actually selected by the corresponding weights in the TPE. Hence, it is not possible such errors to affect the final result. Silent faults increase in the case of 1:4 sparsity compared to the case of 2:4 sparsity. This is the result of faults occurring in the weight registers that remain idle in the case of 1:4 sparsity. False checker reactions cover a small overall percentage that reflects that the probability that a fault hits the checker and not the sparse tensor array is small.

As the number of injected faults per fault-injection campaign increases (1--5 faults are randomly injected) the observed  results change significantly. Fault detection increases above 93\% and the possibility of having a false alarm drops to 1\% on average.

\section{Conclusions}
To reduce inference time and memory storage, ML models are often pruned and quantized.
To allow for online error checking during the execution of such ML models on sparse tensor arrays, we adapt the widely accepted ABFT methodology for online error checking to the unique characteristics of such computation structures. Checksum logic is placed at the periphery of the sparse tensor array and by reusing its operation computes the actual and the predicted checksum of matrix multiplication following the digit-serial arithmetic paradigm. Hardware analysis has validated the low-cost property of the checksum logic, while detailed fault injection campaigns reveals its rich fault detection properties.

\bibliographystyle{IEEEtran}
\bibliography{refs}

\end{document}